\documentclass[prl,aps,twocolumn, floatfix, superscriptaddress,showpacs]{revtex4}
\usepackage{amsmath,bm,graphicx}
\usepackage{amssymb}
\usepackage{amsfonts}
\usepackage{color}
\begin{document}

\title{Running faster together: huge speed up of thermal ratchets due to hydrodynamic coupling \\
}

\author{Paolo Malgaretti}
\email[Corresponding Author : ]{paolomalgaretti@ffn.ub.es }
\affiliation{Department de Fisica Fonamental, Universitat de Barcelona, Spain}

\author{Ignacio Pagonabarraga}
\affiliation{Department de Fisica Fonamental, Universitat de Barcelona, Spain}

\author{Daan Frenkel}
\affiliation{Department of Chemistry, University of Cambridge, Lensfield Road, Cambridge, CB2 1EW, United Kingdom}
 
 \date{\today}

\begin{abstract}
We present simulations that reveal a surprisingly large effect of 
hydrodynamic coupling on the speed of thermal ratchet motors. The model 
that we use considers
particles performing thermal ratchet motion in a hydrodynamic solvent. 
Using particle-based, mesoscopic simulations that maintain local 
momentum conservation, we analyze quantitatively how the coupling to the 
surrounding fluid affects ratchet motion.
We find that coupling can increase the mean velocity of the moving 
particles by almost two orders of magnitude, precisely because ratchet 
motion has both a diffusive and a deterministic component. The resulting 
coupling also leads to the formation of aggregates at longer 
times. The correlated motion that we describe increases the efficiency 
of motor-delivered cargo transport and we speculate that the 
mechanism that we have uncovered may play a key role in speeding up 
molecular motor-driven intracellular transport.
\end{abstract}


\maketitle


The motion of particles in a fluid is affected by the hydrodynamic 
interaction mediated by the embedding medium~\cite{russel}. The dynamic 
coupling of small suspended particles  affects, for instance,  the average 
velocity of particles driven by a constant force~\cite{russel} and the 
correlation spectrum of pairs of freely diffusing 
particles~\cite{Martin2006,Leonardo2008}. All existing studies show that hydrodynamic coupling 
can increase the speed at which particles move, but the resulting speed-ups are typically quite modest~\cite{holger,Houtman2007} .

In this Letter we show that hydrodynamic interactions can cause a very 
large speedup of particles that move asynchronously by a thermal ratchet mechanism: 
hydrodynamic coupling increases the speed of such motors by up to two 
orders of magnitude compared to the velocity of isolated particles. 
Physical realizations of such ratchet motors, to which we will refer generically as steppers, can be created in colloidal 
systems~\cite{Mateos2011} and may be found in molecular motors~\cite{Ajdari1997} that move along polar 
biofilaments, such as microtubules or actin. Hence, the effect of 
hydrodynamic coupling on  stepping particles is likely to be relevant for 
the understanding of the physical mechanisms underlying intracellular 
transport processes such as cytoplasmic streaming~\cite{Goldstein2008}, axonal transport~\cite{Greulich2007,Okada_Hirokawa} and 
membrane-embedded cargo pulling~\cite{Campas2004}.
 
\begin{figure}
\includegraphics[scale = 0.25,angle=-90]{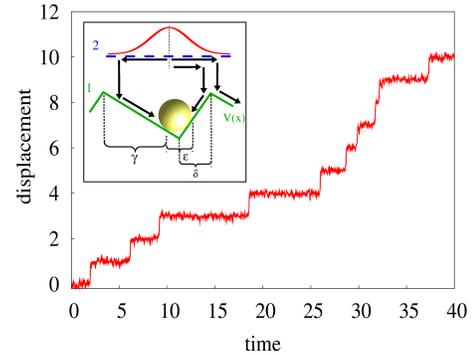}
 \caption{Typical trajectory generated by a two-state ratchet 
 model~\cite{Ajdari1997} for  a single stepper. Inset: free-energy landscape.  In the $\gamma$ region of \emph{state 1} (power stroke phase) the particle experiences the potential $V_1$. In region $\epsilon$ the  particle can jump to the \emph{state 2}  with a rate $p_{12}$. In \emph{state 2}  (diffusive phase) the particle diffuses and can revert to \emph{state 1}  with a rate $p_{21}$. Parameter values used in the simulations: $\gamma=0.75$, $\varepsilon=0.01$, $\delta=0.2$, $\Delta V_1=200$ (maximum energy difference); lengths  in units of the ratchet period $l$ and energy in units of $k_BT$.
}
\label{ratchet-trajectory} 
\end{figure} 
In order to study the behavior of many steppers moving along the same filament, we employ a simple model that accounts both for the essential features of  steppers and for the time-dependent hydrodynamics of the embedding  fluid. The moving particles  are described using the two-state ratchet model~\cite{Ajdari1997},  a standard, simplified model that accounts for the mechanochemical coupling underlying   molecular motor mechanics.  In this ratchet model,  the stepper can be in  two different internal states: in \emph{state 1}  particles displace under the action of a potential, $V(x)$, of period $l$,  which depends on the position, $x$ , along the filament axis and that we consider piecewise-linear  for simplicity. In  \emph{state 2}, particles undergo thermal diffusion along the filament. Steppers can switch from \emph{state 1} to \emph{state 2} with probability $p_{12}$  in a region of width $\epsilon$ around the potential minimum. The time spent diffusing in \emph{state 2} is determined by the homogeneous probability $p_{21}$ to   jump back to \emph{state 1}. For energy consuming particles, the  ratio $p_{12}/p_{21}$  differs from its  thermal equilibrium ratio, thus breaking  detailed balance and leading to rectified motion  along the filament (see Fig.~\ref{ratchet-trajectory})~\footnote{ $p_{12}$ and $p_{21}$ are chosen to optimize  the velocity of  an isolated motor~\cite{Ajdari1997}}.

To account efficiently for hydrodynamics that capture the geometrical confinement in which steppers displace, the fluid  surrounding the  filaments and moving particles is modeled as an ideal dissipative fluid where point particles closer than a prescribed cutoff distance, $r_c$, interact through the Lowe-Andersen thermostat~\cite{Lowe1999}.  Fluid particles  within  $r_c$ from the filament surface or  moving particles exchange momentum  using the same  local thermostat. The model thus enforces  local equilibrium and  linear and angular momentum conservation, yet it does not change the free-energy landscape of the ratchet model.  The spatial extent of  moving particles is taken into account  modeling them as hard, spherical particles with radius $a$. We have studied the collective motion of  steppers that move at a fixed distance $r_0$ from the surface of a straight, fixed, filament  of cross-section $\pi r_0^2$. The filament is  aligned parallel to  one of the edges of our simulation box, of length $L\sim 100\,l$ and square section with edge $L_x=L_y=24\,l$ and periodic boundary conditions are applied  (Similar results have been obtained for cylindrical confinement). For simplicity,   moving particles never detach from the filament (i.e. they   are infinitely processive according to the classification of molecular motors).  Starting from a particle random distribution  along the filament, we  monitor the  stepper average velocity $v$,  and its dispersion on time scales during which  moving particles  can cross the periodic unit box ($t \leq  L/v_0$), being $v_0$ the average velocity of an isolated  particle.

Before turning to  hydrodynamic interactions (HI), we first briefly consider  the role of excluded-volume interactions (EV)  among  steppers.
\begin{figure}
\includegraphics[scale = 0.25,angle=-90]{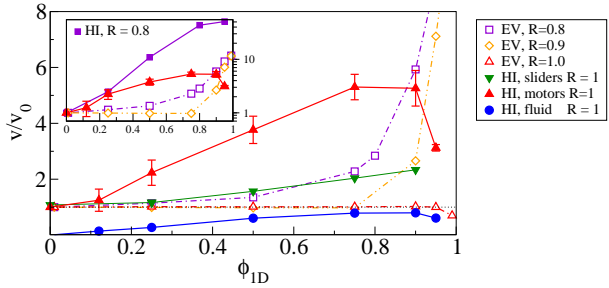}
 \caption{Average velocity  of steppers as a function of particle filament coverage, $\phi_{1D}$.  For comparison the corresponding results for ``sliders'' (see text) are also shown: open and filled down-triangles. 
Inset: comparison between R=1 (triangles) and R=0.8 (squares) steppers including HI.}
\label{no-fluid}
\label{1D-fluid}
\end{figure}
The effect of EV on $v$ depends on  the ratio $R=a/l$. For biological molecular motors $R$ is of order one, but in artificial ratchets it can be tuned by  controlling the range of particle-particle interactions~\cite{u2000}. Figure~\ref{no-fluid} shows the variation of $v$ with the  moving particle  filament coverage. The results in Fig.~\ref{no-fluid} were obtained when particles move along a  single track  (prescribed by a confining angular potential) but this assumption is not essential: similar results are obtained if  particles  are allowed to spread on the filament surface. The mean velocity is sensitive to the precise value of the ratio $R$. A monotonic increase  of  $v$ with filament coverage ($\phi_{1D}=2a N/L$, with  $N$  the number of particles on the filament) is observed for incommensurate   moving particles,  opposed to what is observed for passive diffusion. This behavior can be understood because when the range of  particle-particle repulsion is not commensurate with the ratchet period,  a stepper  in the diffusive state may be pushed  toward  the ratchet maximum (i.e. to the right into $\delta$  in the  inset of Fig.~\ref{ratchet-trajectory}) by a neighbor staying on its left (in region $\gamma$).  Hence, steppers speed up due to the decrease in the time it takes them to move  to the next ratchet minimum.  When the particle size is commensurate with the ratchet period, $R=1$, $v$ depends only weakly on filament coverage, except at very high concentrations where  many-particle ratchet motion becomes less efficient than isolated particle  motion.  The motion of tightly  elastically coupled   motors also displays an analogous dependence on the commensurability of motor separation with  the ratchet period~\cite{Lutz2006, goko_pre_00}.

To focus on the effect of HI, we consider  independent, processive steppers of size $R=1$, where EV are unimportant.
 Fig.~\ref{1D-fluid} shows  the dependence of $v$ for   particles  walking 
 along a filament,  as a function of the fractional filament particle  coverage,  $\phi_{1D}$. For $\bar{\phi}_{1D}\leq 0.1$, the average particle velocity does not depend on concentration but at higher concentrations, we observe a fairly  linear dependence of $v$ on $\phi_{1D}$.  The  velocity increase  is significantly larger than that observed for  hydrodynamically coupled  particles  that move on a surface  under the influence  of a  {\em constant} external force that has been chosen such that it reproduces the average velocity of isolated particles (referred to as``sliders'' in Fig.~\ref{1D-fluid})~\cite{Houtman2007}.  The key difference between sliders and the ratchet motion characteristic of steppers is that sliders move smoothly, rather than in bursts. 
A  simple mean-field argument can be used to estimate the particle density  at which hydrodynamic speed-up becomes significant. The  average fraction  $\rho p_{\downarrow}$  of bound particles  which move  under the  influence of  the ratchet force, $f$, (\emph{state 1}) induces a mean drift velocity  $v \simeq 2\frac{f}{6\pi\eta a}\rho p_{\downarrow}\frac{l-\delta}{l}\intop_{2R}^{L/2}dr\frac{3a}{2r} $ on the diffusing steppers (\emph{state 2}) over the characteristic time $\Delta t \simeq \frac{1}{2}\frac{6\pi\eta a}{f(l-\delta)}$ in which  bound particles displace  along the  filament.  Hydrodynamic correlated motion plays a role when the mean displacement felt by the diffusing particles  due to the ratcheted ones allows them to surmount the characteristic potential barrier, {\sl i.e.}  $v\Delta t \geq \delta$. Hence, hydrodynamic speed-up becomes significant for $\phi_{1D} \geq   \bar{\phi}_{1D}\equiv 2 \delta /(l-d)^2\ln L/4 a$, which corresponds to $\bar{\phi}_{1D}\simeq 0.1$ for $\delta =0.2, R=1, \tilde{L}=100$ and $p_{\downarrow}\sim 1$.

 This hydrodynamic coupling, independent on the direct forces steppers  exert on each other, is qualitatively different from the collective motion of  particles interacting through short range forces where the  possibility to  enhance collectively their velocity depends on the degree of attraction and its commensurability with the potential ratchet~\cite{holger,Brugues2009}. When $\phi_{1D}$ approaches one,  EV lead to  a substantial decrease of  the mean particle  velocity. 
Fig.~\ref{1D-fluid}  also shows that collective motion induces a  net average flow of the fluid in which the filament and particles are embedded.  Hence,  active ratchet  motion favors also the transport of  suspended, passive particles. This transport scenario is likely to be relevant for cargo motion in elongated geometries  found in biological systems  such as neurons and in plant cells. If the particle size is not commensurate with the ratchet period, $R<1$,  the average stepper velocity is further enhanced, as shown in the inset of Fig.~\ref{1D-fluid}. 
\begin{figure}
\includegraphics[scale = 0.25,angle=-90]{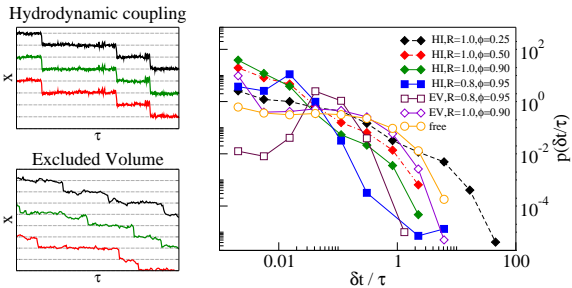}
 \caption{Left: Correlated (uncorrelated) motion of  three adjacent particles with HI (EV). Right: probability distribution of the dwell time, $\delta t$, between the consecutive stepping of neighbor steppers.}
\label{1D-corr-step}
\end{figure}

The probability distribution function (pdf) of  dwell times, $\delta t$, between subsequent stepping events of nearby particles provides insight into the  mechanism of   hydrodynamic velocity enhancement. 
As shown in fig.~\ref{1D-corr-step}, when hydrodynamic coupling is taken into account a sharp peak for small $\delta t$ appears indicating that  HI favors  the correlated motion of subsequent particles. When a particle  steps, it generates a transient flow that pushes (pulls) the particles that are in front (behind) it, thus facilitating their crossing of the ratchet barrier.
A similar speed-up has been described for colloids moving in a saw-tooth potential, under the influence of a constant, external force~\cite{Lutz2006}.  
Such behavior is not observed for EV interactions that, for commensurate motors ($R=1$) show a behavior very close to the, reference, behavior of free motors (nor HI nor EV interaction), while for smaller $R$, the short-ranged interactions between moving particles hinder their relative  displacements,  which favors shorter dwell times. The left panels of Fig.~\ref{1D-corr-step} display  typical  trajectories of three nearby steppers;  in the presence of HI (top), jumps are strongly correlated, while no correlated jumps are observed in the absence of HI (bottom).

\begin{figure}
\includegraphics[scale = 0.25,angle=-90]{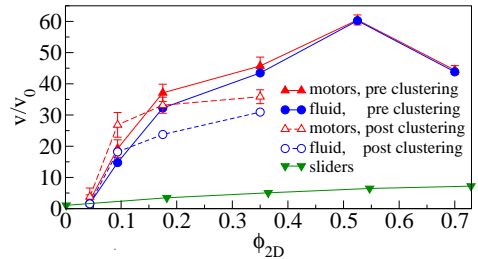}
 \caption{HI-induced clustering of steppers moving on the cylindrical surface of the filament. 
 Filled symbols show the average velocity of steppers (triangles) and fluid (circles) before clustering. The corresponding open symbols show the velocity after clustering. ``Sliders'' (downward triangles)  clustering does not affect the mean velocity.}
\label{2D-fluid}
\end{figure}
Thermal ratchet motors are not necessarily restricted to move along a single  track. For example, in biological systems microtubules indeed are composed by  a number of polarized tracks arranged in a cylindrical filament. To mimic the latter scenario we allowed particles to displace  freely  on the  filament surface while subject to the same filament interaction, $V(x)$. An even stronger increase in $v$ is  observed now as a function of the  surface particle coverage,  $\phi_{2D}= N a^2/(2 r L)$, as shown in Fig.~\ref{2D-fluid}.  The more homogeneous filament  coverage  with respect to  single track motion leads to average particle velocities that are one order of magnitude larger and it results in a stronger coupling to the surrounding fluid; the average flow velocity is comparable to that of the  steppers. This implies that collective active motion can induce efficient  passive intracellular  transport.  As in the single track case, there is a characteristic threshold  coverage, $\bar{\phi}_{2D}$, above which hydrodynamic speed up of $v$ is observed.

\begin{figure}
\includegraphics[scale = 0.27,angle=-90]{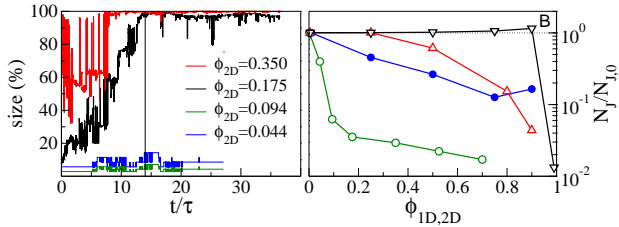}
 \caption{ Left: fraction of steppers belonging to the  largest cluster, as a function of time. The time unit $\tau\equiv L/v_0$. Right: 
Average number of stepper power strokes  to displace one ratchet period. The energy consumption is proportional to the number of power strokes. Up (Down) triangles stands for steppers interacting through EV and R=0.8 (R=1)  while filled (open) circles stand for steppers interacting through HI for single track (free diffusion on the filament surface) motion.}
\label{2D-cluster}
\end{figure}

The fact that the mean particle velocity depends on  the stepper concentration leads to ``bunching''.  A group of moving particles  with a  higher than average  concentration will move faster than the rest, thus leading to the build up of larger clusters. This clustering results in heterogeneous filament coverage. 
We have analyzed  cluster formation of stepper aggregates on long time scales, $t \gg L/v_0$.  Using a robust distance criterion to identify clusters~\footnote{Two steppers belong to the same cluster if their distance along the filament is smaller that $3R$.}, we find that at intermediate and high filament coverage  particles aggregate into a single cluster that survives for the rest of the simulation up to twice the time needed to form the aggregate (see panel A of Fig.~\ref{2D-cluster}). At long times the average particle  speed and the coupling with the embedding fluid decrease slightly, as shown in Fig.~\ref{2D-fluid}, underlining that the heterogeneous coverage of the filament jointly with an increased effective filament coverage caused by clustering affect the particle collective motion. As the mean filament coverage increases, the aggregate percolates along the filament. At low filament coverages, $\phi_{2D}< 0.15$, we do not observe large aggregates. Rather, steppers  organize into small clusters that  form and dissolve. However,  we cannot rule out that clusters  develop at  longer times.  

Clearly, the hydrodynamic speed-up of moving particle groups increases transport  efficiency.  Assuming the transition rate between the two internal  states of a particle  is independent of filament coverage,  the energy consumption of non-interacting steppers is constant.  A good measure of particle displacement efficiency is then the average number of cycles,   $N_J$, between the two internal energy states  required to make a particle move one period.   Panel B of Fig.~\ref{2D-cluster} shows that $N_J$ decreases sharply when the steppers speed-up due to HI.  For EV the decrease in $N_J$ is much more gradual and it is essentially absent if the size of the particle is  commensurate with $V(x)$.   The sharp decrease of $N_J$ with coverage illustrates that, as a result of  hydrodynamic coupling,  many particles ``surf along''  on the flow field generated by the power stroke of a single stepper.\\
 
 We have shown that dynamic interactions among weakly coupled steppers strongly affect their average speed due to the changes in the particles' dwell times  between successive jumps when  the particle motion proceeds in short steps.  Such an interaction leads to a speed-up of up to a factor 60 compared with the velocity of an 
isolated  thermal ratchet motor. Moreover, HI induces a surrounding fluid flow that allows transport of suspended particles with a mean velocity comparable to that of  particles driving the flow. 
  The phenomena that we describe should be relevant and experimentally observable, for example, in the collective transport of colloidal particles moving under the action of a ratchet potential generated by an optical trap although to our knowledge, such experiments have not yet been attempted. We expect our results are also relevant for biological systems   where the steppers are molecular motors pulling a cargo. In particular, the geometry studied  is of particular interest for axonal transport~\cite{Greulich2007,Okada_Hirokawa} where the processive motor KIF1A has been proposed as a good experimental realization of the two state ratchet model. More generally, we expect our results to be relevant whenever the heterogeneous dynamic described by the two state model is fulfilled. This  is found , for example,  of non-processive motors whose tails are embedded in a membrane-covered cargo. In this case the fluid-like membrane~\cite{Campas2004} will sustain the hydrodynamic coupling while the proximity of the cargo 
to the filament will enhance the binding rate of non-processive motors. 
Finally we speculate that the strong velocity enhancement that we 
observe may play a role in the very fast cytoplasmic streaming that has 
been observed in some plant cells~\cite{Goldstein2008}.

  P.M. and I.P. acknowledge financial support from MINECO  (Spain) and DURSI under projects  FIS2011-
22603  and 2009SGR-634, respectively. D.F. acknowledges support from ERC Advanced Grant  227758,  Wolfson Merit Award 2007/R3 of the Royal Society of London and EPSRC Programme Grant EP/I001352/1.

\bibliography{letter2}

\begin{thebibliography}{15}
\expandafter\ifx\csname natexlab\endcsname\relax\def\natexlab#1{#1}\fi
\expandafter\ifx\csname bibnamefont\endcsname\relax
  \def\bibnamefont#1{#1}\fi
\expandafter\ifx\csname bibfnamefont\endcsname\relax
  \def\bibfnamefont#1{#1}\fi
\expandafter\ifx\csname citenamefont\endcsname\relax
  \def\citenamefont#1{#1}\fi
\expandafter\ifx\csname url\endcsname\relax
  \def\url#1{\texttt{#1}}\fi
\expandafter\ifx\csname urlprefix\endcsname\relax\def\urlprefix{URL }\fi
\providecommand{\bibinfo}[2]{#2}
\providecommand{\eprint}[2][]{\url{#2}}

\bibitem[{\citenamefont{Russel et~al.}(1992)\citenamefont{Russel, Saville, and
  Schowalter}}]{russel}
\bibinfo{author}{\bibfnamefont{W.~B.} \bibnamefont{Russel}},
  \bibinfo{author}{\bibfnamefont{D.~A.} \bibnamefont{Saville}},
  \bibnamefont{and} \bibinfo{author}{\bibfnamefont{W.~R.}
  \bibnamefont{Schowalter}}, \emph{\bibinfo{title}{Colloidal Dispersions}}
  (\bibinfo{publisher}{Cambridge University Press},
  \bibinfo{address}{Cambridge}, \bibinfo{year}{1992}).

\bibitem[{\citenamefont{Martin et~al.}(2006)\citenamefont{Martin, Reichert,
  Stark, and Gisler}}]{Martin2006}
\bibinfo{author}{\bibfnamefont{S.}~\bibnamefont{Martin}},
  \bibinfo{author}{\bibfnamefont{M.}~\bibnamefont{Reichert}},
  \bibinfo{author}{\bibfnamefont{H.}~\bibnamefont{Stark}}, \bibnamefont{and}
  \bibinfo{author}{\bibfnamefont{T.}~\bibnamefont{Gisler}},
  \bibinfo{journal}{PRL} \textbf{\bibinfo{volume}{97}}, \bibinfo{pages}{248301}
  (\bibinfo{year}{2006}).

\bibitem[{\citenamefont{Leonardo et~al.}(2008)\citenamefont{Leonardo, Keen,
  Ianni, Leach, Padgett, and Ruocco}}]{Leonardo2008}
\bibinfo{author}{\bibfnamefont{R.~D.} \bibnamefont{Leonardo}},
  \bibinfo{author}{\bibfnamefont{S.}~\bibnamefont{Keen}},
  \bibinfo{author}{\bibfnamefont{F.}~\bibnamefont{Ianni}},
  \bibinfo{author}{\bibfnamefont{J.}~\bibnamefont{Leach}},
  \bibinfo{author}{\bibfnamefont{M.~J.} \bibnamefont{Padgett}},
  \bibnamefont{and} \bibinfo{author}{\bibfnamefont{G.}~\bibnamefont{Ruocco}},
  pp. \bibinfo{pages}{1--4} (\bibinfo{year}{2008}).

\bibitem[{\citenamefont{Grimm and Stark}(2011)}]{holger}
\bibinfo{author}{\bibfnamefont{A.}~\bibnamefont{Grimm}} \bibnamefont{and}
  \bibinfo{author}{\bibfnamefont{H.}~\bibnamefont{Stark}},
  \bibinfo{journal}{Soft Matter} \textbf{\bibinfo{volume}{7}},
  \bibinfo{pages}{3219} (\bibinfo{year}{2011}).

\bibitem[{\citenamefont{Houtman et~al.}(2007)\citenamefont{Houtman,
  Pagonabarraga, Lowe, Emons, and Eiser}}]{Houtman2007}
\bibinfo{author}{\bibfnamefont{D.}~\bibnamefont{Houtman}},
  \bibinfo{author}{\bibfnamefont{I.}~\bibnamefont{Pagonabarraga}},
  \bibinfo{author}{\bibfnamefont{C.~P.} \bibnamefont{Lowe}},
  \bibinfo{author}{\bibfnamefont{A.~M.~C.} \bibnamefont{Emons}},
  \bibnamefont{and} \bibinfo{author}{\bibfnamefont{E.}~\bibnamefont{Eiser}},
  \bibinfo{journal}{EPL} \textbf{\bibinfo{volume}{78}}, \bibinfo{pages}{10}
  (\bibinfo{year}{2007}).

\bibitem[{\citenamefont{Mateos et~al.}(2011)\citenamefont{Mateos, Arzola, and
  Volke-sepu}}]{Mateos2011}
\bibinfo{author}{\bibfnamefont{L.}~\bibnamefont{Mateos}},
  \bibinfo{author}{\bibfnamefont{A.~V.} \bibnamefont{Arzola}},
  \bibnamefont{and}
  \bibinfo{author}{\bibfnamefont{K.}~\bibnamefont{Volke-sepu}},
  \bibinfo{journal}{Physical Review Letters} p. \bibinfo{pages}{168104}
  (\bibinfo{year}{2011}).

\bibitem[{\citenamefont{Julicher et~al.}(1997)\citenamefont{Julicher, Ajdari,
  and Prost}}]{Ajdari1997}
\bibinfo{author}{\bibfnamefont{F.}~\bibnamefont{Julicher}},
  \bibinfo{author}{\bibfnamefont{A.}~\bibnamefont{Ajdari}}, \bibnamefont{and}
  \bibinfo{author}{\bibfnamefont{J.}~\bibnamefont{Prost}},
  \bibinfo{journal}{Rev. Mod. Phys.} \textbf{\bibinfo{volume}{69}},
  \bibinfo{pages}{1269} (\bibinfo{year}{1997}).

\bibitem[{\citenamefont{Goldstein et~al.}(2008)\citenamefont{Goldstein, Tuval,
  Meent, and W}}]{Goldstein2008}
\bibinfo{author}{\bibfnamefont{R.~E.} \bibnamefont{Goldstein}},
  \bibinfo{author}{\bibfnamefont{I.}~\bibnamefont{Tuval}},
  \bibinfo{author}{\bibfnamefont{V.~D.} \bibnamefont{Meent}}, \bibnamefont{and}
  \bibinfo{author}{\bibfnamefont{J.}~\bibnamefont{W}}, \bibinfo{journal}{Proc.
  Natl. Acad. Sci. USA} \textbf{\bibinfo{volume}{105}}, \bibinfo{pages}{3663}
  (\bibinfo{year}{2008}).

\bibitem[{\citenamefont{Greulich et~al.}(2007)\citenamefont{Greulich, Garai,
  Nishinari, Schadschneider, and Chowdhury}}]{Greulich2007}
\bibinfo{author}{\bibfnamefont{P.}~\bibnamefont{Greulich}},
  \bibinfo{author}{\bibfnamefont{A.}~\bibnamefont{Garai}},
  \bibinfo{author}{\bibfnamefont{K.}~\bibnamefont{Nishinari}},
  \bibinfo{author}{\bibfnamefont{A.}~\bibnamefont{Schadschneider}},
  \bibnamefont{and}
  \bibinfo{author}{\bibfnamefont{D.}~\bibnamefont{Chowdhury}},
  \bibinfo{journal}{PRE} \textbf{\bibinfo{volume}{75}}, \bibinfo{pages}{041905}
  (\bibinfo{year}{2007}).

\bibitem[{\citenamefont{Okada and Hirokawa}(1999)}]{Okada_Hirokawa}
\bibinfo{author}{\bibfnamefont{Y.}~\bibnamefont{Okada}} \bibnamefont{and}
  \bibinfo{author}{\bibfnamefont{N.}~\bibnamefont{Hirokawa}},
  \bibinfo{journal}{Science} \textbf{\bibinfo{volume}{283}},
  \bibinfo{pages}{1152} (\bibinfo{year}{1999}).

\bibitem[{\citenamefont{Campas et~al.}(2004)\citenamefont{Campas, Zeldovich,
  Jolimaitre, Bourel-Bonnet, Bassereau, Prost, and Goud}}]{Campas2004}
\bibinfo{author}{\bibfnamefont{O.}~\bibnamefont{Campas}},
  \bibinfo{author}{\bibfnamefont{K.~B.} \bibnamefont{Zeldovich}},
  \bibinfo{author}{\bibfnamefont{P.}~\bibnamefont{Jolimaitre}},
  \bibinfo{author}{\bibfnamefont{L.}~\bibnamefont{Bourel-Bonnet}},
  \bibinfo{author}{\bibfnamefont{P.}~\bibnamefont{Bassereau}},
  \bibinfo{author}{\bibfnamefont{J.}~\bibnamefont{Prost}}, \bibnamefont{and}
  \bibinfo{author}{\bibfnamefont{B.}~\bibnamefont{Goud}},
  \bibinfo{journal}{Proc. Natl. Acad. Sci. USA} \textbf{\bibinfo{volume}{101}},
  \bibinfo{pages}{17096} (\bibinfo{year}{2004}).

\bibitem[{\citenamefont{Lowe}(1999)}]{Lowe1999}
\bibinfo{author}{\bibfnamefont{C.~P.} \bibnamefont{Lowe}},
  \bibinfo{journal}{Europhysics Letters} \textbf{\bibinfo{volume}{47}},
  \bibinfo{pages}{145} (\bibinfo{year}{1999}).

\bibitem[{\citenamefont{Pelton}(2000)}]{u2000}
\bibinfo{author}{\bibfnamefont{R.}~\bibnamefont{Pelton}}, \bibinfo{journal}{Adv
  Coll Int Sci} \textbf{\bibinfo{volume}{85}}, \bibinfo{pages}{1}
  (\bibinfo{year}{2000}).

\bibitem[{\citenamefont{Lutz et~al.}(2006)\citenamefont{Lutz, Reichert, Stark,
  and Bechinger}}]{Lutz2006}
\bibinfo{author}{\bibfnamefont{C.}~\bibnamefont{Lutz}},
  \bibinfo{author}{\bibfnamefont{M.}~\bibnamefont{Reichert}},
  \bibinfo{author}{\bibfnamefont{H.}~\bibnamefont{Stark}}, \bibnamefont{and}
  \bibinfo{author}{\bibfnamefont{C.}~\bibnamefont{Bechinger}},
  \bibinfo{journal}{Europhysics Letters} \textbf{\bibinfo{volume}{74}},
  \bibinfo{pages}{719} (\bibinfo{year}{2006}).

\bibitem[{\citenamefont{Brugues and Casademunt}(2009)}]{Brugues2009}
\bibinfo{author}{\bibfnamefont{J.}~\bibnamefont{Brugues}} \bibnamefont{and}
  \bibinfo{author}{\bibfnamefont{J.}~\bibnamefont{Casademunt}},
  \bibinfo{journal}{PRL} \textbf{\bibinfo{volume}{102}},
  \bibinfo{pages}{118104} (\bibinfo{year}{2009}).

\end{thebibliography}

\end{document}